\documentclass{ws-procs9x6}

\newcommand{\ie}{{\it i.e.}}
\newcommand{\eg}{{\it e.g.}}

\begin{document}

\title{GRAD-SHAFRANOV APPROACH TO
AXISYMMETRIC STATIONARY FLOWS IN ASTROPHYSICS}

\author{V.S. Beskin}
\address{Lebedev Physical Institute, Leninskii prosp., 53, Moscow, 119991,
Russia. E-mail: beskin@lpi.ru}

\maketitle

\abstracts{ My lecture is devoted to the analytical results
available for a large class of axisymmetric stationary flows in
the vicinity of compact astrophysical objects. First, the most
general case is formulated corresponding to the axisymmetric
stationary MHD flow in the Kerr metric. Then, I discuss the
hydrodynamical version of the Grad-Shafranov equation. Although
not so well-known as the full MHD one, it allows us to clarify the
nontrivial structure of the Grad-Shafranov approach as well as to
discuss the simplest version of the $3+1$-split language -- the
most convenient one for the description of ideal flows in the
vicinity of rotating black holes.  Finally, I consider several
examples that demonstrate how this approach can be used to obtain
the quantitative description of the real transonic flows in the
vicinity of rotating and moving black holes. }

Many astrophysical sources are axisymmetric and stationary to a
good accuracy. These include both accreting neutron stars and
black holes, axisymmetric stellar (solar) winds, jets from young
stellar objects, and ejection of particles from magnetospheres of
rotating neutron stars. It can not be ruled out that such
magnetohydrodynamic flows also play an important role in other
galactic sources, \eg, microquasars. The latter are regarded as
candidates for black holes not to say about active galactic nuclei
where the electrodynamical processes in the vicinity of the
rotating supermassive black holes are considered as the most
reasonable model of their central engine~\cite{1}$^{,}$\cite{2}.
So, it is not surprising that ideal magnetohydrodynamics, which
allows sufficiently simple formalization of the problem, is
actively applied when describing these flows.

The point is that due to axial symmetry and stationarity (as well
as the ideal freezing-in condition), in the most general case it
is possible to introduce five integral of motions which are
constant at axisymmetric magnetic surfaces. This remarkable fact
allows us to separate the problem of finding the poloidal field
structure (the poloidal flow structure in the hydrodynamics) from
the problem of particle acceleration and the structure of electric
currents. The solution of the latter task for a given poloidal
field can be obtained in terms of quite simple algebraic
relations. It is important that such an approach can be
straightforwardly generalized to flows in the vicinity of the
rotating black holes, as the Kerr metric is also axially symmetric
and stationary.

On the other hand, it is much more difficult to find the
two-dimensional poloidal magnetic field structure (the
hydrodynamical flow structure). First of all, this is due to the
complex structure of the equation describing axisymmetric
stationary flows. In the general case, it is a nonlinear equation
of the mixed type, which changes from elliptical to hyperbolical
at singular surfaces and in addition contains integrals of motion
in the form of free functions. Generally speaking, similar
equations, which stem from the classical Tricomi equation, have
been discussed since the beginning of the last century in
connection with transonic hydrodynamic flows~\cite{3}. Later on,
the equations describing axially symmetric stationary flows were
called Grad-Shafranov equations after the authors who formulated
in the late 1950s an equation of such a type in connection with
controlled thermonuclear fusion~\cite{4}. This equation, however,
was originally related to equilibrium static configurations only
and required strong revision when it was generalized to the
transonic case. The full version of such an equation was
formulated by L.S. Soloviev in 1963 in the third volume of {\it
Problems of Plasma Theory}~\cite{5} and was well-known to
physicists. However, as it often occurs, the full version of the
Grad-Shafranov equation was little known in the astrophysical
literature, so it was later `rediscovered' scores of
times~\cite{6}.

As it turned out, the difficulty lay in the fact that the very
formulation of the direct problem within the Grad-Shafranov
approach proved to be nontrivial. For example, in the
hydrodynamical limit, when there are only three integrals of
motion, the problem requires four boundary conditions for the
transonic flow regime. This implies that, for instance, two
thermodynamic functions and two velocity components should be
specified at some surface. However, to determine the Bernoulli
integral, which naturally should be known in order to solve the
Grad-Shafranov equation, all three components of the velocity must
be specified, which is impossible since the third velocity
component itself is to be obtained from the solution. This is in
fact one of the main difficulties of the approach under
consideration.

Nevertheless, several approaches exist that allow us to construct
the analytical solution of direct problems within the framework of
the Grad-Shafranov method. For example, this is possible when the
exact solution of this equation is known and we explore flows
weakly diverging from the known one. Spherically symmetric
accretion (ejection) of matter could be such an exact solution. As
a result, the known structure of the flow in the zeroth
approximation enables us to determine (with the required accuracy)
both  the location of singular surfaces and all the integrals of
motion directly from boundary conditions, thus making it possible
to solve the
 Grad-Shafranov equation within the direct formulation of a problem.

\section{Grad-Shafranov equation}

Let us consider the axisymmetric stationary plasma flow in the
vicinity of a rotating black hole, \ie, in the Kerr
metric~\cite{2}:
\begin{equation}
{\rm d}s^{2}=-\alpha^{2}{\rm d}t^{2} +g_{ik}({\rm
d}x^{i}+\beta^{i}{\rm d}t)({\rm d}x^{k}+\beta^{k}{\rm d}t),
\label{a1}
\end{equation}
where
\begin{eqnarray}
 & & \alpha=\frac{\rho}{\Sigma}\sqrt{\Delta},\qquad \beta^{r}=\beta^{\theta}=0,
 \qquad \beta^{\varphi}=-\omega=-\frac{2aMr}{\Sigma^{2}}, \nonumber \\
 & & g_{rr}=\frac{\rho^{2}}{\Delta},\qquad  g_{\theta \theta}=\rho^{2},
 \qquad g_{\varphi \varphi}=\varpi^{2}.
\label{a2}
\end{eqnarray}
Here $\alpha$ is the lapse function (gravitational red shift)
vanishing on the horizon
\begin{equation}
r_{\rm g} = M +\sqrt{M^2-a^2},
\end{equation}
$\omega$ is the angular velocity of local nonrotating observers
(the so-called Lense-Thirring angular velocity), and
\begin{eqnarray}
 & & \Delta=r^{2}+a^{2}-2 Mr, \qquad \rho^{2}=r^{2}+a^{2}\cos^{2}\theta,
 \nonumber \\
 & & \Sigma^{2}=(r^{2}+a^{2})^{2}-a^{2}\Delta\sin^{2}\theta, \qquad
 \varpi=\frac{\Sigma}{\rho}\sin\theta.
\label{a4}
\end{eqnarray}
As usual, $M$ and $a$ are the black hole mass and angular momentum
per unit mass ($a = J/M$) respectively. Here indices without hats
denote components of vectors with respect to the coordinate basis
$\partial/\partial r$, $\partial/\partial \theta$, and
$\partial/\partial \varphi$, and indices with hats correspond to
their physical components. Finally, below we shall use the system
of units with $c = G = 1$.

In what follows we shall also use the $3+1$ split
language~\cite{2}. Within this approach, the physical quantities
are expressed in terms of three-dimensional vectors which would be
measured by observers moving around the rotating black hole with
the angular velocity $\omega$ (so-called ZAMOs -- zero angular
momentum observers). The convenience of the $3+1$ split language
is connected with the fact that the representation of many
expressions has the same form as in the flat space. On the other
hand, all thermodynamic quantities are determined in the comoving
reference frame.

Now, we shall demonstrate how the five `integrals of motions',
which are constant at the magnetic surfaces, can be derived in the
general case of axisymmetric stationary flows. It is convenient to
introduce the scalar function $\Psi(r,\theta)$ which has a meaning
of magnetic flux. As a consequence, the magnetic field is defined
in the following way:
\begin{equation}
{\bf B} = \frac{{\bf\nabla}\Psi \times {\bf e}_{\hat
\varphi}}{2\pi\varpi} -\frac{2I}{\alpha\varpi}{\bf e}_{\hat
\varphi}, \label{a21}
\end{equation}
where $I(r,\theta)$ is the total electric current inside the
region $\Psi<\Psi(r,\theta)$.

As usual, we assume that the magnetosphere contains sufficient
amount of plasma to satisfy the freezing-in condition which, using
the $3+1$ split language, preserves the form ${\bf E}+{\bf
v}\times{\bf B} = 0$. On the other hand, the stationarity (as well
as the condition for zero longitudinal electric field) implies
that the field ${\bf E}$ can be written as
\begin{equation}
{\bf E}=-\frac{\Omega_{\rm F}-\omega}{2\pi\alpha}{\bf\nabla}\Psi.
\label{a24}
\end{equation}
By substituting relation (\ref{a24}) into the Maxwell equations,
it is easy to verify that the condition ${\bf B} \cdot \nabla
\Omega_{\rm F} = 0$ is satisfied, \ie\ that $\Omega_{\rm F}$ must
be constant at the magnetic surfaces (Ferraro's isorotation law):
\begin{equation}
\Omega_{\rm F} = \Omega_{\rm F}(\Psi).
\end{equation}

Next, the Maxwell equation $\nabla \cdot{\bf B} = 0$, the
continuity equation, and the freezing-in condition allow us to
write the four-velocity of matter ${\bf u}$ in the form
\begin{equation}
{\bf u}=\frac{\eta}{\alpha n}{\bf B}
+\gamma(\Omega_F-\omega)\frac{\varpi} {\alpha}{\bf
e}_{\hat\varphi}, \label{a26}
\end{equation}
where $\gamma=1/\sqrt{1-v^2}$ is the Lorentz factor of matter
(measured by ZAMOs), and the quantity $\eta$ is the particle flux
to magnetic flux ratio. Due to the relationship ${\bf\nabla}\cdot
(\eta{\bf B}_{p})=0$, $\eta$ must be constant at the magnetic
surfaces $\Psi(r,\theta) =$ const as well, \ie,
\begin{equation}
\eta = \eta(\Psi).
\end{equation}

The next two integrals of motions result from our assumption that
the flow is axisymmetric and stationary. This yields the
conservation law of the energy $E$ and the $z$-component of
angular momentum $L_{z}$:
\begin{eqnarray}
 & & E=E(\Psi)=\frac{\Omega_{\rm F}I}{2\pi}+\mu\eta(\alpha\gamma+\omega u_
{\varphi});
\label{a31} \\
 & & L=L(\Psi)=\frac{I}{2\pi}+\mu\eta \varpi u_{\hat\varphi},
\label{a32}
\end{eqnarray}
where $\mu=(\rho_{m}+P)/n$ is the relativistic enthalpy
($\rho_{m}$ is the internal energy density, and $P$ is the
pressure). Finally, in the axially symmetric case the isentropy
condition yields
\begin{equation}
s=s(\Psi),
\end{equation}
so that the entropy per particle, $s(\Psi)$, is the fifth integral
of motion.

The five integrals of motions $\Omega_{\rm F}(\Psi)$,
$\eta(\Psi)$, $s(\Psi)$, $E(\Psi)$, and $L(\Psi)$, as well as the
poloidal magnetic field ${\bf B}_{\rm p}$, allow us to find the
toroidal magnetic field $B_{\hat \varphi}$ and all other plasma
parameters:
\begin{eqnarray}
\frac{I}{2\pi} & = & \frac{\alpha^{2}L-(\Omega_{\rm
F}-\omega)\varpi^{2} (E-\omega L)}{\alpha^{2}-(\Omega_{\rm
F}-\omega)^{2}\varpi^{2}-{\bf M}^2};
\label{a33} \\
\nonumber \\
\gamma & = & \frac{1}{\alpha\mu\eta}\frac{\alpha^{2}(E-\Omega_{\rm
F}L)-{\bf M}^2 (E-\omega L)}{\alpha^{2}-(\Omega_{\rm
F}-\omega)^{2}\varpi^{2}-{\bf M}^2};
\label{a34} \\
\nonumber \\
u_{\hat\varphi} & = & \frac{1}{\varpi\mu\eta}\frac{(E-\Omega_{\rm
F}L)
 (\Omega_{\rm F}-\omega)\varpi^{2}-L{\bf M}^2}{\alpha^{2}-(\Omega_{\rm F}
 -\omega)^{2}\varpi^{2}-{\bf M}^2},
\label{a35}
\end{eqnarray}
where
\begin{equation}
{\bf M}^2=\frac{4\pi\eta^{2}\mu}{n}. \label{a36}
\end{equation}
It is easy to see that ${\bf M}^2$ is proportional (with the
factor of $\alpha^2$) to the Mach number squared of the poloidal
velocity $u_{\rm p}$ with respect to the Alfv\'en velocity $u_{\rm
A}=B_{\rm p}/\sqrt{4\pi n\mu}$, \ie, ${\bf M}^2=\alpha^{2}u_{\rm
p}^2/u_{\rm A}^2$.

Since $\mu = \mu(n,s)$, definition (\ref{a36}) allows us to
express the concentration $n$ (and hence the specific enthalpy
$\mu$) as a function of $\eta$, $s$, and ${\bf M}^2$. This means
that along with the five integrals of motion, the expressions for
$I$, $\gamma$, and $u_{\hat\varphi}$ depend only on one additional
quantity, namely the Mach number ${\bf M}$. To determine the Mach
number ${\bf M}$, it is necessary to use the obvious relation
$\gamma^{2}-{\bf u}^{2}=1$, which, owing to equations (\ref{a34})
and (\ref{a35}), can be rewritten in the form
\begin{equation}
\frac{K}{\varpi^{2}A^{2}}=\frac{1}{64\pi^{4}}\frac{{\bf
M}^4({\bf\nabla} \Psi)^{2}}{\varpi^{2}}+\alpha^{2}\eta^{2}\mu^{2},
\label{a38}
\end{equation}
where
\begin{equation}
A=\alpha^{2}-(\Omega_{\rm F}-\omega)^{2}\varpi^{2}-{\bf M}^2
\label{a39}
\end{equation}
and
\begin{eqnarray}
K=\alpha^{2}\varpi^{2}(E-\Omega_{\rm F}L)^{2}
\left[\alpha^{2}-(\Omega_{\rm F}- \omega)^{2}\varpi^{2}-2{\bf
M}^2\right]
\nonumber \\
+{\bf M}^4\left[\varpi^{2}(E-\omega L)^{2}-\alpha^{2}L^{2}\right].
\label{a40}
\end{eqnarray}

As for the Grad-Shafranov equation itself, \ie, the equilibrium
equation for magnetic field lines, it can be written in the form
\begin{eqnarray}
& & \frac{1}{\alpha}\nabla_{k}\left\{\frac{1}{\alpha\varpi^2}
[\alpha^{2}-(\Omega_{\rm F}-\omega)^{2}\varpi^{2}-{\bf
M}^2]\nabla^{k}\Psi\right\} +\frac{\Omega_{\rm F}
-\omega}{\alpha^{2}}({\bf\nabla}\Psi)^{2}\frac{{\rm d} \Omega_{\rm
F}}{{\rm d}\Psi}
\nonumber \\
\label{a64}\\
& & +\frac{64\pi^{4}}{\alpha^{2}\varpi^{2}}\frac{1}{2{\bf M}^2}
\frac{\partial}{\partial\Psi}\left(\frac{G}{A}\right)
-16\pi^{3}\mu n\frac{1}{\eta}\frac{{\rm d}\eta}{{\rm d}\Psi}
-16\pi^{3}nT\frac{{\rm d}s}{{\rm d}\Psi}=0,  \nonumber
\end{eqnarray}
where
\begin{equation}
G=\alpha^{2}\varpi^{2}(E-\Omega_{\rm F}L)^{2}+\alpha^{2}{\bf
M}^2L^{2}-{\bf M}^2 \varpi^{2}(E-\omega L)^{2}, \label{a65}
\end{equation}
and the derivative $\partial/\partial\Psi$ acts on the integrals
of motion only. Finally, expressing in Eq. (\ref{a64}) the terms
$\nabla_{k}{\bf M}^{2}$ according to Eq. (\ref{a40}) we
obtain~\cite{7}
\begin{eqnarray}
 & & A\left[\frac{1}{\alpha}\nabla_{k}\left(\frac{1}{\alpha\varpi^{2}}
 \nabla^{k}\Psi\right)+\frac{1}{\alpha^{2}\varpi^{2}({\bf\nabla}\Psi)^{2}}
 \frac{\nabla^{i}\Psi\cdot\nabla^{k}\Psi\cdot\nabla_{i}\nabla_{k}\Psi}{D}\right]
 \nonumber\\
 & & +\frac{\nabla_{k}'A\nabla^{k}\Psi}{\alpha^{2}\varpi^{2}}
 -\frac{A}{\alpha^{2}\varpi^{2}({\bf\nabla}\Psi)^{2}}\frac{1}{2D}
 \nabla_{k}'F\nabla^{k}\Psi+\frac{\Omega_{\rm F}-\omega}{\alpha^{2}}
 ({\bf\nabla}\Psi)^{2}\frac{{\rm d}\Omega_{\rm F}}{{\rm d}\Psi}
 \label{main}\\
 & & + \frac{64\pi^{4}}{\alpha^{2}\varpi^{2}}\frac{1}{2{\bf M}^{2}}
 \frac{\partial}
 {\partial\Psi}\left(\frac{G}{A}\right)-16\pi^{3}\mu n\frac{1}{\eta}
 \frac{{\rm d}\eta}{{\rm d}\Psi}-16\pi^{3}nT\frac{{\rm d}s}{{\rm d}\Psi}=0. \nonumber
\end{eqnarray}
Here
\begin{equation}
D=\frac{A}{{\bf M}^{2}}+\frac{\alpha^{2}}{{\bf M}^{2}}
\frac{B^{2}_{\hat\varphi}}{B^{2}_{\rm p}}-\frac{1}{u^{2}_{\rm p}}
\frac{A}{{\bf M}^{2}}\frac{c^{2}_{\rm s}}{1-c^{2}_{\rm s}},
\end{equation}
\begin{equation}
F=\frac{64\pi^{4}}{{\bf M}^{4}}\frac{K}{A^{2}}
-\frac{64\pi^{4}}{{\bf M}^{4}}\alpha^{2}
\varpi^{2}\eta^{2}\mu^{2},
\end{equation}
and the gradient $\nabla_{k}'$ denotes the action of $\nabla_{k}$
under the condition that ${\bf M}$ is fixed. Let us stress that in
equation (\ref{main}) the pressure $P$, the temperature $T$, sound
velocity $c_{\rm s}$, and the specific enthalpy $\mu$ are to be
expressed via an equation of state in terms of the entropy
$s(\Psi)$ and the square of the Mach number ${\bf M}^{2}$. In
turn, the quantity ${\bf M}^{2}$ is to be considered as a function
of $({\bf\nabla}\Psi)^{2}$ and the integrals of motion,
\begin{equation}
{\bf M}^{2}={\bf M}^{2}\left[({\bf\nabla}\Psi)^{2}, E(\Psi),
L(\Psi), \eta(\Psi), \Omega_{\rm F}(\Psi), s(\Psi)\right].
\end{equation}
The latter relation is the implicit form of Eq.\ (\ref{a38}). The
stream equation~(\ref{main}) coupled with definitions
(\ref{a21})~--~(\ref{a32}) is the desired equation for the
poloidal field which contains only the magnetic flux $\Psi$ and
the five integrals of motion $\Omega_{\rm F}(\Psi), \eta(\Psi),
s(\Psi), E(\Psi)$, and $L(\Psi)$ depending on it.

Equation (\ref{main}) is a second-order equation linear with
respect to the highest derivatives. It changes its type from
elliptical to hyperbolical at singular surfaces where the poloidal
velocity of matter becomes equal to either fast or slow
magnetosonic velocity ($D = 0$), or to the cusp velocity ($D =
-1$). Although at the Alfv\'enic surface, $A = 0$, the type of
equation does not change, the Alfv\'enic surface does represent a
singular surface of the Grad-Shafranov equation because a
regularity condition must be satisfied there.

\section{Examples}

{\bf Bondi-Hoyle accretion}. As a first example, we consider the
hydrodynamic accretion onto a moving back hole (the Bondi-Hoyle
accretion), which is one of the classical problems of modern
astrophysics~\cite{2}. First of all, let us formulate the
hydrodynamical limit of the Grad-Shafranov equation, where we can
neglect the electromagnetic field contribution. In this case, it
is convenient to introduce a new potential $\Phi(\Psi)$ satisfying
the condition $\eta(\Psi)={\rm d}\Phi/{\rm d}\Psi$. Using
definition (\ref{a26}) we obtain
\begin{equation}
\alpha n{\bf u}_{\rm p}=
\frac{1}{2\pi\varpi}({\bf\nabla}\Phi\times{\bf e} _{\hat\varphi}).
\label{b2}
\end{equation}
Surfaces $\Phi(r, \theta)=$ const define the streamlines of
matter.

In the hydrodynamic limit, there are only three integrals of
motion. These are the energy flux and the $z$-component of the
angular momentum:
\begin{eqnarray}
 & & E(\Phi)=\mu (\alpha\gamma+\varpi \omega u_{\hat\varphi});
\label{b3} \\
 & & L(\Phi)=\mu\varpi u_{\hat\varphi},
\label{b4}
\end{eqnarray}
as well as the entropy $s=s(\Phi)$. Now the algebraic Bernoulli
equation (\ref{a38}) takes the form
\begin{equation}
(E-\omega L)^{2}=\alpha^{2}\mu^{2}+\frac{\alpha^2}{\varpi^2}L^2+
\frac{\hat {\bf M}^4}{64\pi^{4}\varpi^2}(\nabla\Phi)^2, \label{b6}
\end{equation}
where the `Mach number' squared $\hat {\bf M}^2$ is defined as
$\hat {\bf M}^2=4\pi\mu/n$. Then the Grad-Shafranov equation
(\ref{a64}) can be rewritten in the form~\cite{7}
\begin{eqnarray}
&&-\frac{1}{\alpha}\nabla_{k} \left(\frac{\hat {\bf
M}^2}{\alpha\varpi^{2}}\nabla^{k}\Phi\right)
-16\pi^{3}nT\frac{{\rm d}s}{{\rm d}\Phi}
\nonumber \\
&&+\frac{64\pi^{4}}{\alpha^{2}\varpi^{2}\hat {\bf M}^2}
\left[\varpi^{2}(E-\omega L)\left(\frac{{\rm d}E}{{\rm d}\Phi}
-\omega\frac{{\rm d}L}{{\rm d}\Phi}\right) -\alpha^{2}L\frac{{\rm
d}L}{{\rm d}\Phi}\right] =0, \label{b8}
\end{eqnarray}
where now
\begin{equation}
D = -1+\frac{1}{u^{2}_{\rm p}}\frac{c^{2}_{s}}{1-c^{2}_{s}}.
\label{b9}
\end{equation}
As we see, equation (\ref{b8}) contains only one singular surface,
\ie\ the sonic surface, determined from the condition $D = 0$.

To construct the solution corresponding to the Bondi-Hoyle
accretion, it is possible to seek the solution of the
Grad-Shafranov equation for the flux function $\Phi(r,\theta)$ in
the form of a small perturbation of the spherically symmetric flow
in the reference frame moving with the black hole
\begin{equation}
\Phi(r,\theta )=\Phi_{0}[1 - \cos\theta + \varepsilon_1
f(r,\theta)]. \label{b31}
\end{equation}
Here we introduce a small parameter
\begin{equation}
\varepsilon_1 = \frac{v_{\infty}}{c_{\infty}}
\end{equation}
which defines the ratio of the black hole velocity to the velocity
of sound at infinity. For a nonmoving gravity center we return to
the spherically symmetric flow.

As Grad-Shafranov equation (\ref{b8}) contains three invariants,
it is necessary to specify four boundary conditions, say
\begin{enumerate}
\item $v_{{\rm p},\infty} =$ const, \item $v_{\varphi} = 0$ (and
hence $L = 0$), \item $s_{\infty} =$ const, \item $E_{\infty} =
c_{\infty}^2/(\Gamma - 1)$.
\end{enumerate}
In the last relation we neglect the terms $\sim \varepsilon_1^2$.

As a result, the Grad-Shafranov equation can be linearized:
\begin{equation}
-\varepsilon_1 \alpha^2 D\frac{\partial^{2} f}{\partial r^2}
-\frac{\varepsilon_1}{r^2}(D+1)\sin\theta
\frac{\partial}{\partial\theta}\left(\frac{1}{\sin\theta}
\frac{\partial f}{\partial\theta}\right) +\varepsilon_1 \alpha^2
N_r \frac{\partial f}{\partial r} = 0, \label{b32}
\end{equation}
where
\begin{equation}
N_r = \frac{2}{r} - \frac{\mu^2}{E^2 - \alpha^2\mu^2} \,
\frac{M}{r^2}.
\end{equation}
It is extremely important that according to (\ref{b2}) and
(\ref{b9}),
\begin{equation}
D + 1 = \frac{\alpha^2\mu^2}{E^2-\alpha^2\mu^2}\cdot \frac{c_{\rm
s}^2}{1-c_{\rm s}^2}, \label{c49}
\end{equation}
so the factor $\alpha^2$ enters every term of equation
(\ref{b32}). Hence, equation (\ref{b32}) has no singularity at the
horizon. In particular, it means that it is not necessary to
specify any boundary  conditions for $r = r_{\rm g}$. It is not
surprising because the horizon corresponds to the supersonic
region which cannot affect the subsonic flow.

On the other hand, we see that all the terms contain the small
value $\varepsilon_1$. Hence, the functions $D$, $c_{\rm s}$, {\it
etc.} can be taken from the zeroth solution. As for the
spherically symmetric flow, the functions $D$, $c_{\rm s}$, {\it
etc.} do not depend on $\theta$, and the solution of equation
(\ref{b32}) can be expanded in eigen functions of the operator
$\sin\theta \,
\partial/\partial\theta (1/\sin\theta \cdot
\partial/\partial\theta)$. Thus, the solution can be presented in
the form
\begin{equation}
f(r,\theta) = \sum_{m=0}^{\infty} g_m(r) Q_m(\theta),
\end{equation}
the equations for the radial functions $g_m(r)$ being
\begin{equation}
r^2 D\frac{{\rm d}^{2} g_m}{{\rm d} r^2} + r^2 N_r \frac{{\rm d}
g_m}{{\rm d} r} +m(m+1) \, \frac{\mu^2}{E^2-\alpha^2\mu^2} \,
\frac{c_{\rm s}^2}{1-c_{\rm s}^2} \, g_m = 0. \label{b34}
\end{equation}
Here $Q_0 = 1 -\cos\theta$, $Q_1 = \sin^2\theta$, $Q_2 =
\sin^2\theta\cos\theta$, $\dots$ are the eigen functions of the
angular operator.

As to the boundary conditions, they can be formulated as follows:
\begin{enumerate}
\item No singularity on the sonic surface (where $N_r = 0$, $D =
0$),
\begin{equation}
g_m(r_{*}) = 0. \label{b35}
\end{equation}
\item The homogeneous flow $\Phi = \pi n_{\infty}v_{\infty}
r^2\sin^2\theta$ at infinity which gives
\begin{equation}
g_1 \rightarrow
\frac{1}{2}\,\frac{n_{\infty}c_{\infty}}{n_{*}c_{*}}
\,\frac{r^2}{r_{*}^2}, \qquad g_2, g_3, \dots = 0. \label{b36}
\end{equation}
\end{enumerate}
As a result, the complete solution can be presented in the form
\begin{equation}
\Phi(r,\theta) = \Phi_{0}[1 - \cos\theta + \varepsilon_1
g_1(r)\sin^2\theta)], \label{b37}
\end{equation}
where the radial function $g_1(r)$ is the solution of the ordinary
differential equation (\ref{b34}) for $m = 1$ with the boundary
conditions (\ref{b35}) and (\ref{b36}).

\begin{figure}[htbp]
        \centerline{\psfig{file=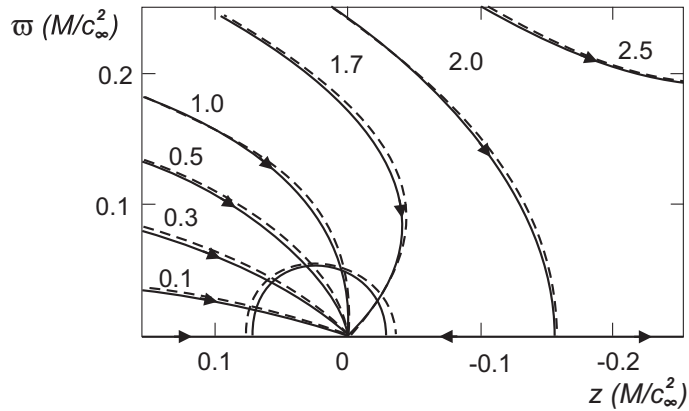,width=9cm}}
        \caption{Bondi-Hoyle accretion onto moving black hole.}
\end{figure}

This means that we have constructed the analytical solution of the
problem, \ie, obtained the full description of the flow structure.
For example, the sonic surface has now the nonspherical form
\begin{equation}
r_{*}(\theta) = r_{*}\left[1 + \varepsilon_1
\left(\frac{\Gamma+1}{5 - 3\Gamma}\right)k_2 \cos\theta\right],
\label{b38}
\end{equation}
where the numerical coefficient $k_2 = r_{*} g_1^{\prime}(r_{*})$
is expressed through the derivative of the radial function
$g_1(r)$ at the sonic point. As shown in Fig.~1, the analytical
solution fully agrees with the numerical calculations~\cite{hunt}
in spite of the parameter $\varepsilon_1 = 0.6$ here being quite
large. Here $\Gamma = 4/3$, and the numbers alongside the curves
denote values $\Phi/\Phi_0$; the dashed lines show the streamlines
and the form of the sonic surface obtained numerically.

As can be easily seen, outside the capture radius
\begin{equation}
R_{\rm c} \approx \varepsilon_1^{-1/2} r_*
\end{equation}
our main assumption, \ie, the smallness of the deviation from the
spherically symmetric flow, is  not valid. Nevertheless, the
solution found remains correct. This remarkable property is due to
the Grad-Shafranov equation becoming linear for constant
concentration $n$. But as we learn from the spherically symmetric
Bondi accretion, at large distances $r \gg r_*$ from the sonic
surface the density of the accreting matter is virtually constant.
Accordingly, the concentration is constant for a homogeneous flow
as well. As a result, under the condition that $R_{\rm c} \gg
r_*$, which holds true for $\varepsilon_1 \ll 1$, near and beyond
the capture radius (where the perturbation $\sim
\varepsilon_1g_1(r)$ becomes comparable to unity) equation
(\ref{b8}) becomes linear. So that the sum of the two solutions,
homogeneous and spherically symmetric ones, is also a solution.

{\bf Thin transonic disk}. As a next example, we consider the
internal two-dimensional structure of a thin accretion disk. Here,
for simplicity we consider the case of a nonrotating
(Schwarzschild) black hole~\cite{btchekh}. We recall that
according to the standard model~\cite{s1} the accreting matter
forms an equilibrium disk rotating around the gravitational center
with the Keplerian velocity $v_{\rm K}(r) =
\left(GM/r\right)^{1/2}$. The disk will be thin provided that its
temperature is sufficiently small ($c_{\rm s} \ll v_{\rm K}$)
since the vertical balance of the gravity force and the pressure
gradient implies that
\begin{equation}
H \approx r \frac{c_{\rm s}}{v_{\rm K}}.
\end{equation}
The General Relativity effects result in two important properties:
the absence of stable circular orbits for $r < r_0 = 3r_{\rm g}$
and the transonic regime of accretion. The first point means that
the accreting matter passing the marginally stable orbit
approaches the black hole horizon sufficiently fast, namely, in
the dynamical time $\tau_{\rm d} \sim [v_r(r_0)/c]^{-1/3}r_{\rm
g}/c$. It is important that such a flow is realized in the absence
of viscosity. The second statement results from the fact that up
to the marginally stable orbit the flow is subsonic while at the
horizon the flow is to be supersonic.

It is necessary to stress that the existence of the small
parameter $\varepsilon_2 = u_0/c_0 \ll 1$, where $c_0$ is the
 sound velocity and $u_0$ is the gas radial velocity on
the marginally stable orbit, comes from the relation $v_r/v_{ \rm
K} \approx \alpha_{\rm ss} c_{\rm s}^2/v_{\rm K}^2$ for the radial
velocity in the accretion disk~\cite{1}. In the vicinity of the
marginally stable orbit this estimate is apparently inapplicable.
Nevertheless, below we shall consider the parameter
$\varepsilon_{2}$ to be small because the presence of a small
parameter allows us to investigate the flow structure
analytically. In addition, the small parameter makes the effect
under discussion more visible.

Up to now in the majority of works devoted to thin accretion disks
the procedure of vertical averaging was used, where the vertical
four-velocity $u_{\hat \theta}$ was assumed to be
zero~\cite{pbk81}. As a result, the vertical component of the
dynamic force $nu^{b}\nabla_{b}(\mu u_{a})$ in the Euler equation
was postulated to be inimportant up to horizon. For this reason
the disk thickness was determined by the pressure gradient even in
the supersonic region near the black hole~\cite{alp}. Here I am
going to demonstrate that the assumption $u_{\hat \theta} = 0$ is
not correct. As for the Bondi accretion, the dynamic force is to
be important in the vicinity of the sonic surface.

\begin{figure}[htbp]
        \centerline{\psfig{file=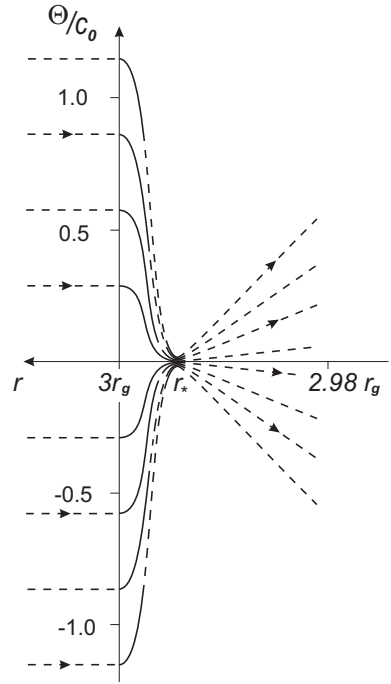,width=9cm}}
      \caption{Nozzle shape of a thin transonic disk near marginally
      stable orbit.}
\end{figure}

Figure 2 shows the structure of a thin accretion disk after
passing the marginally stable orbit $r = 3r_{\rm g}$ obtained by
solving equation (\ref{b8}) numerically for $c_0 = 10^{-2}$, $u_0
= 10^{-5}$, \ie, in the presence of a small parameter
$\varepsilon_2 = u_0/c_0 \ll 1$. The solid lines correspond to the
range of parameters $u_{\rm p}^2/c_0^2 < 0.2$. The dashed lines
indicate the extrapolation of the solution to the sonic-surface
region. In the vicinity of the sonic surface the flow has the form
of the standard nozzle.

As we see, the flow structure near the sonic surface is far from
being radial. The appearance of the narrow waist has a simple
physical meaning. Indeed, the density remains almost constant for
subsonic flow while the radial velocity increases from $u_0$ to
$c_* \sim c_0$, \ie, for $\varepsilon_{2} \ll 1$ changes over
several orders of magnitude. As a result, the disk thickness $H$
should change in the same proportion owing to the continuity
equation as well,
\begin{equation}
H(r_*) \approx \frac{u_0}{c_0}H(3r_{\rm g}). \label{compr}
\end{equation}
Here it is extremely important that both components of the
dynamical force become comparable with the pressure gradient near
the sonic surface:
\begin{equation}
\frac{u_{\hat \theta}}{r}\, \frac{\partial u_{\hat
\theta}}{\partial \theta} \approx u_{\hat r}\frac{\partial u_{\hat
\theta}}{\partial r} \approx \frac{\nabla_{\hat \theta}P}{\mu}
\approx \frac{c_0^2}{u_0^2}\, \frac{\theta}{r},
\end{equation}
where the angle $\theta$ is counted off from the equatorial plane.

In other words, if there appears a nonzero vertical velocity
component, the dynamical term $({\bf v}\nabla){\bf v}$ cannot be
neglected in the vertical force balance near the sonic
surface~\cite{btchekh}. It is clear that this property remains
valid for arbitrary radial velocity of the flow, \ie, even when
the transverse contraction of the disk is not so pronounced.
Taking dynamical forces into account causes two additional degrees
of freedom to appear, which relate to the higher derivatives in
the Grad-Shafranov equation. This also leads to extra conclusions
independent of the value of $\varepsilon_{2}$. In thin accretion
disks, the critical condition at the sonic surface does not fix
the accretion rate any more; it determines the bending of the
streamlines near the sonic surface.

Finally, the inclusion of the vertical velocity inevitably leads
to the appearance of a small longitudinal scale $\delta
r_{\parallel} \approx H_*$ in the vicinity of the sonic surface,
which for a thin disk proves to be much smaller than the distance
to the black hole for any value of the parameter $u_0/c_0$. In the
standard one-dimensional approach, this scale does not emerge. As
for the supersonic region (and, in particular, the region near the
horizon), the disk thickness here will be determined not by the
pressure gradient, but by the form of ballistic trajectories.

{\bf The Blandford-Znajek process}. In conclusion, we discuss the
energy loss of a rotating black hole embedded in an external
magnetic field -- the so-called Blandford-Znajek
process~\cite{bz}. This process is considered to be the most
preferential mechanism of energy release in active galactic
nuclei, microquasars, and even cosmological gamma-ray
bursters~\cite{lwb00}. Its main idea is based on the analogy with
the energy transfer in the internal regions of radio pulsar
magnetospheres. Indeed, let us suppose that there is a regular
external magnetic field in the vicinity of a rotating black hole,
and that the electric current $I$ flows along magnetic field
lines. Then, the electric field ${\bf E}$, which is induced by
plasma rotating with the angular velocity  $\Omega_{\rm F}$, and
the toroidal magnetic field ${\bf B}_{\varphi}$, which is due to
the longitudinal current $I$, generate the electromagnetic energy
flux (the Poynting vector flux) carrying the energy away along the
magnetic field lines.

Of course, by definition, general relativity effects are important
near the black hole. Consequently, it is not obvious that the
pulsar analogy can be useful in all cases. For example, in radio
pulsars, the braking of neutron stars results from the Amp\'ere
force acting on the star surface. This force results from the
surface currents shorting the electric currents flowing in the
pulsar magnetosphere~\cite{ufn1}. In the case of black holes such
currents cannot lead to deceleration as the event horizon is not a
physically preferred surface, though surface currents themselves
can be formally introduced in the framework of the so-called
membrane paradigm~\cite{2}.

Indeed, let us consider the well-known condition at the horizon
(the absence of infinite electromagnetic fields in the reference
frame comoving with freely falling observer) $E_{\theta}^{\prime}
\rightarrow (E_{\theta} + B_{\varphi})/\alpha < \infty$. It means
that $E_{\theta} + B_{\varphi} \rightarrow 0$ which can be
rewritten in the form
\begin{equation}
4\pi I(\Psi) =\left[\Omega_{\rm H}-\Omega_{\rm F}(\Psi)\right]
\frac{r_ {\rm g}^2+a^2}{r_{\rm g}^2+a^2\cos^2\theta}\sin\theta
\left(\frac{{\rm d}\Psi}{{\rm d}\theta}\right)_{r_{\rm g}}.
\label{d6}
\end{equation}
Here $a = J/M$ is the rotation parameter. This relation was
used~\cite{bz} as the `boundary condition on the horizon'. But it
is clear that the horizon is not causally connected with the outer
space. For this reason the conclusion was made that there is no
energy flux along magnetic field lines passing through the black
hole horizon~\cite{punsly}, and hence a black hole cannot work as
a unipolar inductor extracting the rotation energy by
electromagnetic stresses.

However, a recent more accurate analysis~\cite{bk00} (in which, in
fact, the first solution of the Grad-Shafranov equation for
nonzero particle mass in the Kerr metric was obtained) indicates
that the braking torque acts in the plasma generation region above
the black hole horizon. Such a torque appears due to the action of
long-range gravitomagnetic forces which penetrate into the regions
causally connected with the outer space. As a result, it was shown
that for a finite mass of particles in the very vicinity of the
horizon, there is a hyperbolic region of the Grad-Shafranov
equation which is altogether absent in the force-free
approximation. Hence, the full version of the Grad-Shafranov
equation needs no boundary condition on the horizon. Actually,
this property was already demonstrated above by the example of the
Bondi-Hoyle accretion. Thus, in this case it is impossible to
consider relation (\ref{d6}) as a boundary condition. Relation
(\ref{d6}) is automatically true for any solution of the
Grad-Shafranov equation which can be extended up to the horizon.

On the other hand, in the force-free approximation, when the
Grad-Shafranov equation remains elliptical up to the black hole
horizon, extra condition (\ref{d6}) is to be included into
consideration~\cite{uzd}. But as was demonstrated~\cite{ufn}, this
condition is actually the manifestation of the critical condition
on the fast magnetosonic surface. Hence, in reality this condition
is specified on the surface which does not coincide with the event
horizon, and, hence, is in the causal connection with the outer
magnetosphere.

Thus, the Blandford-Znajek mechanism of electromagnetic energy
extraction from rotating black holes faces no causality problem.
As in the pulsar magnetosphere, if there is enough secondary
plasma to screen the longitudinal electric field, its charge
density and electric currents produce the flux of electromagnetic
energy propagating from the central star to infinity. For the same
reason, a rotating black hole embedded into an external magnetic
field works as a unipolar inductor extracting its energy of
rotation by the flux of the electromagnetic energy. As a result,
the energy loss can be evaluated as $W_{\rm tot} \approx W_{\rm
BZ}$, where
\begin{eqnarray}
W_{\rm BZ} = \frac{\Omega_{\rm F}(\Omega_{\rm H} - \Omega_{\rm
F})}{\Omega_{\rm H}^2} \left(\frac{a}M\right)^2 B_0^{2}r_{\rm g}^2
c
\nonumber  \\
\approx 10^{45} \left(\frac{a}M\right)^2
\left(\frac{B_0}{10^{4}\,{\rm G}}\right)^2 \left(\frac {M}{10^9
M_{\odot}}\right)^2 {\rm erg/s}. \label{d13'}
\end{eqnarray}
It is easy to check that for the extreme rotation of a black hole
($a$ = 1) and for $B = B_{\rm Edd} \approx 10^4 (M/{10^9 {\bf}
M}_{\odot})^{-1/2}$ G, the energy loss $W_{\rm BZ}$ (\ref{d13'})
coincides with the Eddington luminosity.

\begin{figure}[htbp]
        \centerline{\psfig{file=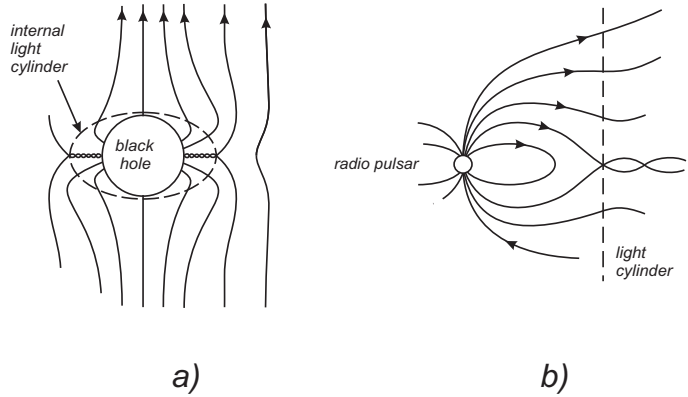,width=9cm}}
        \caption{Magnetospheres of rotating black hole (a)
        and radio pulsar (b).}
\label{fig3}
\end{figure}

It should be emphasized, however, that as follows from equation
(\ref{d13'}), the rate of the energy release needed to explain the
luminosity of active galactic nuclei can be achieved only for the
extreme black hole mass $\sim 10^9 M_{\odot}$, the extreme
magnetic field $B \sim B_{\rm Edd}$, and the extreme angular
velocity $a \sim M$. Therefore, some papers have recently appeared
in which the efficiency of the Blandford-Znajek process in real
astrophysical conditions was questioned~\cite{lop99}. In
particular, it was pointed out that for rapid rotation the Wald
solution for the vacuum magnetosphere leads to the magnetic field
being pushed out from the horizon into the ergosphere~\cite{2},
which causes the appearance of the additional factor $(1 -
a^2/M^2) \rightarrow 0$ in expression (\ref{d13'}).

But as shown in Fig.~\ref{fig3}a, in the black hole magnetosphere
filled with plasma, all magnetic field lines crossing the surface
of the internal light cylinder
$\alpha^{2}=(\Omega_{F}-\omega)^{2}\varpi^{2} + {\bf M}^2$ are to
cross the black hole horizon as well, that is why to an order of
magnitude the energy loss for extremely rotating black holes
coincides with the loss given by equation (\ref{d13'}). Indeed,
here the situation is to be fully analogous to the pulsar
magnetosphere where the field lines passing the external light
cylinder do not intersect the equatorial plane (Fig.~\ref{fig3}b).
The magnetic field structure shown in Fig.~\ref{fig3}a was
recently obtained numerically~\cite{koide}.

\section{Conclusion}

In some simple cases the Grad-Shafranov equation allows us to
construct the exact solution to the problem. In particular, this
approach is very useful in studying the analytical properties of
transonic flows and in determining the required number of boundary
conditions. On the other hand, in the general case no consistent
procedure exists regarding the construction of the solution within
the Grad-Shafranov approach. The point is that the location of
singular surfaces, at which critical conditions should be
formulated, is not known beforehand and itself must be found from
the solution to the problem. Moreover, it is impossible to
generalize this approach to the case of nonideal, non-axially
symmetric and nonsteady flows. So it is not surprising that most
investigators, who are in the first place interested in
astrophysical applications, have focused on a totally different
class of equations, namely on time relaxation problems, that can
only be solved numerically~\cite{wils}. Nevertheless, it is clear
that the key physical results obtained using the Grad-Shafranov
approach are independent of the computing method. For this reason
one can hope that the results presented above can be useful for
everyone working in this field.

\section*{Acknowledgments}

The author is greatly indebted to Professor Seok Jae Park for the
hospitality and also thanks Prof. Hyun Kyu Lee and Hongsu Kim for
fruitful discussions. This work is supported by Grant N
02--02--16762 by Russian Foundation for Basic Researches.

\end{document}